\documentclass[a4paper,11pt]{article}
\pdfoutput=1 % if your are submitting a pdflatex (i.e. if you have
\usepackage{jinstpub} % for details on the use of the package, please
\usepackage{lineno}
\usepackage[utf8]{inputenc} % allow utf-8 input
\usepackage[T1]{fontenc}    % use 8-bit T1 fonts
\usepackage{hyperref}       % hyperlinks
\usepackage{url}            % simple URL typesetting
\usepackage{booktabs}       % professional-quality tables
\usepackage{amsfonts}       % blackboard math symbols
\usepackage{nicefrac}       % compact symbols for 1/2, etc.
\usepackage{microtype}      % microtypography
\usepackage{lipsum}
\usepackage{graphicx}
\usepackage{amsmath}
\usepackage{algorithm}
\usepackage{listings}
\usepackage{setspace}

\lstdefinestyle{mystyle}{
    basicstyle=\ttfamily\footnotesize,
    breakatwhitespace=false,         
    breaklines=true,                 
    captionpos=b,                    
    keepspaces=true,                                 
    showspaces=false,                
    showstringspaces=false,
    showtabs=false,                  
    tabsize=2
}

\usepackage[noend]{algpseudocode}
\usepackage[usenames,dvipsnames,table,modulo]{xcolor}
\usepackage{caption}
\usepackage{subcaption}

\title{A goodness-of-fit test based on a recursive product of spacings}

\author[1,a,b]{Philipp Eller \note{philipp.eller@tum.de}}
\author[2,c]{and Lolian Shtembari \note{lolian@mpp.mpg.de}}
\affiliation[a]{Technical University Munich, Garching, Germany}
\affiliation[b]{Exzellenzcluster ORIGINS, Garching, Germany}
\affiliation[c]{Max Planck Institute for Physics, Munich, Germany}

\arxivnumber{2111.02252} % only if you have one

\keywords{test statistic, hypothesis test, goodness of fit, spacing, interarrival time, waiting time, gap}

\abstract{We introduce a new statistical test based on the observed spacings of ordered data.
The statistic is sensitive to detect non-uniformity in random samples, or short-lived features in event time series.
Under some conditions, this new test can outperform existing ones, such as the well known Kolmogorov-Smirnov or Anderson-Darling tests, in particular when the number of samples is small and differences occur over a small quantile of the null hypothesis distribution.  
A detailed description of the test statistic is provided including a detailed discussion of the parameterization of its distribution via asymptotic bootstrapping as well as a novel per-quantile error estimation of the empirical distribution.
Two example applications are provided, using the test to boost the sensitivity in generic "bump hunting", and employing the test to detect supernovae.
The article is rounded off with an extended performance comparison to other, established goodness-of-fit tests.
}

\begin{document}
\maketitle
\flushbottom

\section{Introduction}

Assessing the goodness-of-fit of a distribution given a number of random samples is an often-encountered problem in data analysis. Such statistical hypothesis tests find applications in many fields, ranging from the natural and social sciences over engineering to quality control.
Several non-parametric tests exist, some of which have become standard tools, including the Kolgogorov-Smirnov (KS) test \citep{Kolmogorov, Smirnov1948TableFE} or the Anderson-Darling (AD) test \citep{AD}.
\cite{Marhuenda} provides a comprehensive overview of existing tests, and a comparison of their performance for the case of detecting non-uniformity for a set of alternative distributions.

In this work, we are in contrast interested in the case where the bulk of samples are actually distributed according to the null hypothesis, and only few additional samples are introduced that are following a different distribution, representing a narrow excess over a known background.
We present the new test statistic "recursive product of spacings", or short RPS, that is based on the spacings between ordered samples, and introduced in Sec.~\ref{sec:rps}.
In Sec.~\ref{sec:cdf} we provide a parametrization of its distribution based on simulations, introducing techniques to estimate the asymptotic result of infinite bootstrapping steps in order to improve the quality of our fits. Subsequently we discuss the quality of the approximation deriving a per-quantile error estimate up to a desired confidence level.
The rest of the article focuses on some illustrations and example applications, as well as a detailed performance comparison to several other test statistics.

\subsection{Goodness-of-fit Tests}

Suppose that we have obtained $n$ samples $x_i$, and want to quantitatively test the hypothesis of those samples being random variates of a known distribution $f(x)$, i.e. independent and identically distributed (i.i.d.) according to $f(x)$. Here, we consider only continuous distributions $f(x)$ with cumulative $F(x)$, and hence can transform samples onto the unit interval 
$[0,1]$ via $y_i = F(x_i)$. This reduces the task at hand to test transformed samples $y_i$ being distributed according to the standard uniform distribution $\mathcal{U}(0,1)$.

First, let us briefly introduce other, existing test statistics to which we will compare the RPS statistic. We consider in particular two groups of statistics, those based on the empirical distribution (EDF Statistics), and those based on the spacings between ordered samples (Spacings Statistics).
An comprehensive overview of existing test statistics can be found in \cite{Marhuenda}.

\subsubsection{EDF Statistics}

This class of test statistics compares the empirical distribution function (EDF) $F_n(x)$ to the cumulative distribution function (CDF) $F(x)$, (here $F(x) = x$). In particular the following are widely used and considered here:

\begin{itemize}
    \item Kolmogorov-Smirnov (KS) \citep{Kolmogorov, Smirnov1948TableFE}: $D_n= \sup_x |F_n(x)-F(x)|$
    \item Cramer-von-Mises (CvM) \citep{cramer, VonMises}: $T = n \int_{-\infty}^\infty (F_n(x) - F(x))^2 dF(x)$
    \item Anderson-Darling (AD) \citep{AD}: $A^2 = n \int_{-\infty}^\infty  \frac{(F_n(x) - F(x))^2}{F(x)\; (1-F(x))} \, dF(x)$
\end{itemize}

Similar are a type of statistics defined on the ordered set. Given the $n$ samples $\{x_1, x_2, \dots , x_n\}$, we define the ordered set of samples as $\{x_{(1)}, x_{(2)}, \dots, x_{(n)}\}$, where $x_{(i)} < x_{(i+1)}\, \forall i$.
The expected value of ordered sample $i$ is $i/(n+1)$, and we define the deviation to the expected values as $\delta_i = x_{(i)} - i/(n+1)$ for each sample $i$. Based on this we can write out the following two statistics:

\begin{itemize}
    \item Pyke's Modified KS (C) \citep{Pyke, Durbin}: $C_n = \max ( \max (\delta_i), -\min(\delta_i))$
    \item Brunk's Modified KS (K) \citep{Brunk}: $K_n = \max (\delta_i) - \min(\delta_i)$
\end{itemize}

\subsubsection{Spacings Statistics}

Based on the ordered set, we can further define the $n+1$ spacings $s$ as $s_i = x_{(i)} - x_{(i-1)}$, with $x_{(0)} = 0$ and $x_{(n+1)}=1$. Several test statistics built from these spacings are considered in literature, including:
\begin{itemize}
    \item Moran (M) \citep{10.2307/2336673}: $M = -\sum_{i=1}^{n+1}\log s_i$
    \item Greenwood (G) \citep{Greenwood}: $G = \sum_{i=1}^{n+1} s^2_i$
\end{itemize}

In the context of a fixed rate Poisson process, these spacings can also be interpreted as \textit{interarrival times} or \textit{waiting times}. In some other areas, spacings are also referred to as \textit{gaps}. 

So-called \textit{higher order spacings} can be defined by summing up neighbouring spacings. Here we consider the overlapping $m$-th order spacings $s_i^{(m)}= x_{(i+m)} - x_{(i)}$. With those, we can define generalisations of Moran and Greenwood, respectively, as discussed by Cressie:

\begin{itemize}
    \item Logarithms of higher order spacings (Lm) \citep{CressieL}: $L_n^{(m)} = -\sum_{i=0}^{n-m+1}\log s_i^{(m)}$
    \item Squares of higher order spacings (Sm) \citep{CressieS}: $S_n^{(m)} = \sum_{i=0}^{n-m+1} (s_i^{(m)})^2$
\end{itemize}

For our comparisons presented later, we choose $m=2$ and $m=3$, respectively, to limit ourselves to a finite list of tests.

Other statistics based on spacings exist and are being actively developed and used, such as, for example, tests based on the $k$ smallest or largest spacings \citep{shtembari2020sum}.

\section{Recursive Product of Spacings (RPS)}
\label{sec:rps}

In this work, our goal is to construct a new test statistic, that is sensitive to narrow features or clusters in an otherwise uniform distribution of samples. The tell-tale sign we are looking for is a localized group of uncommonly small spacings of the ordered data.
For this purpose, we propose a new class of test statistics, that are including higher order spacings in a recursive way.

The recursive product of spacings (RPS) can be thought of as an extension of the Moran statistic, and is defined as

\begin{equation}
    RPS(n) = M^{n+1} + M^n + \dots + M^1,
\end{equation}

where the term $M^{n+1}$ is the \textit{simple} sum of negative log spacings equivalent to the Moran statistic

\begin{equation}
    M^{n+1} = -\sum_{i=1}^{n+1}\log\left( s_i \right).
\label{M_n_+1}
\end{equation}

\begin{figure}[H]
    \centering
    \includegraphics[width=0.6\textwidth]{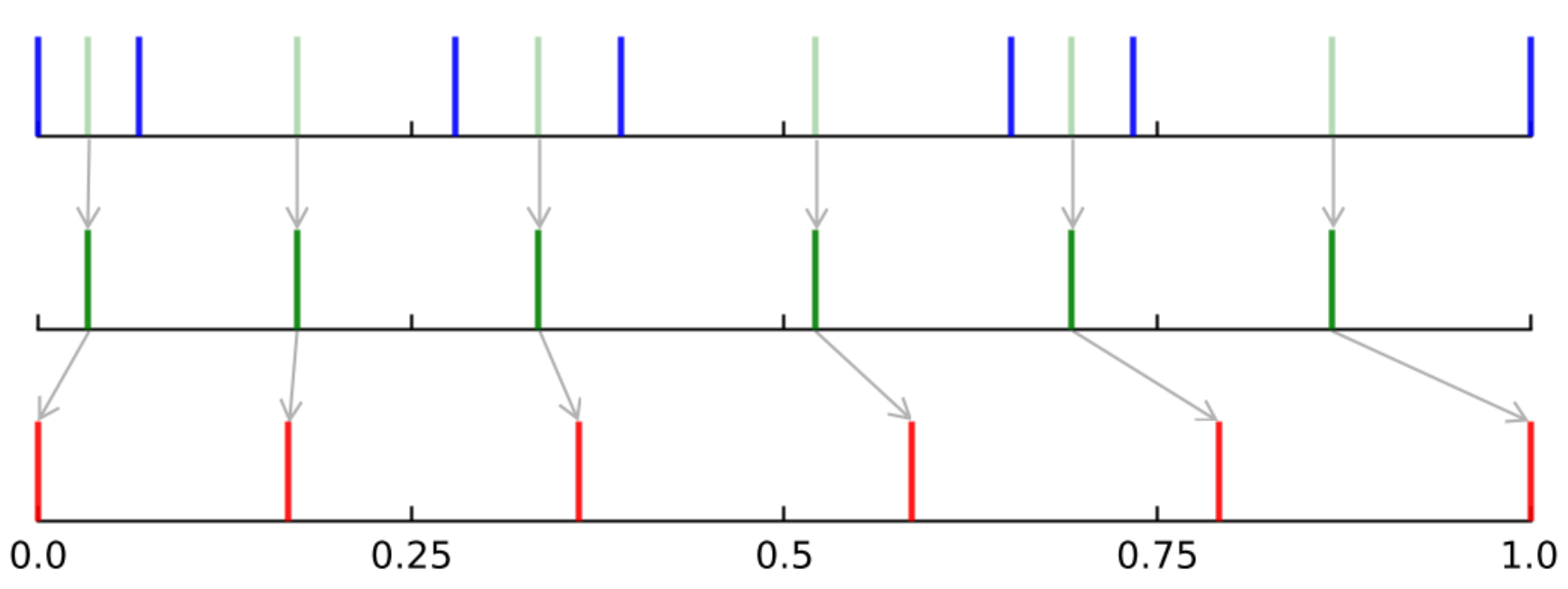}
    \caption{Example of the reduction step included in the $RPS$ calculation. Given an initial set of events (top; blue), the middle points are calculated (top and middle; green), which are then scaled in order to fill the $[0,1]$ interval, forming a new set of data (bottom; red). The evolution of sample positions on the $[0,1]$ interval are annotated via the arrows.}
    \label{fig:reduction_example}
\end{figure}

All following terms are computed in the same way

\begin{equation}
    M^{j} = -\sum_{i=1}^{j}\log\left( s_i^j \right),
\label{M_j}
\end{equation}

but with modified spacings $s_i^j$, defined as:
\begin{equation}
s^j_i = \frac{s^{j+1}_{i} + s^{j+1}_{i+1}}{\sum_i s^j_i}
\label{sij}
\end{equation}

which there are $j$ of, and that depend on the spacings $s^{j+1}$ used to compute the previous term $M^{j+1}$ (hence the \textit{recursiveness}). In order to better understand Eq.~\ref{sij} we can turn to Fig.~\ref{fig:reduction_example}, where we show how to transition from $s^{j+1}$ (top) to $s^j$ (bottom): in the top plot we show a list of events (blue), where we also highlight the boundaries 0 and 1 since they contribute to defining spacings; in the middle plot the middle points of the top row spacings are shown, forming a reduced set of "events", which is then transformed in order to ensure that the spacings of the new set sum up to 1, as shown in the bottom plot; the number of spacings going from the top plot to the bottom one is reduced by one, showing how we have a finite number of reduction steps in the definition of the $RPS$.

We can see that term $M^n$ is identical to $L_n^{(2)}$ up to a normalization factor $1/{\sum_i s_i}$. It is important to include such a normalization for the spacings of each level, as this ensures that the case of equidistant spacings---the most regular and uniform case---yields the smallest possible RPS value.
This minimum value of $RPS(n)$, given by the configuration of equidistant samples, can be expressed easily, as each spacing $s^j$ is equal to $\frac{1}{j}$, and thus:

\begin{equation}
    \min(RPS(n)) = -\sum_{j=1}^{n+1} \sum_{i=1}^{j} \log\left( \frac{1}{j} \right) = \sum_{j=i}^{n+1} j \cdot \log(j).
\label{min_RPS}
\end{equation}

At the other extreme, very small spacings will yield a large contribution to the sum of Eq.~\ref{M_j}, thus $\max(RPS(n))= \infty$ for any given number of samples $n$.
These extrema show that RPS measures the irregularity in sample positions. The RPS statistic increases the more samples aggregate into local clusters.

The RPS quantity calculated so far has an infinite support.
Since approximating the tails of distributions is often easier when dealing with a bounded quantity instead, we transform the $RPS$ in order to bound its support in the range $[0,1]$, via

\begin{equation}
    RPS^*(n) = \frac{\min(RPS(n))}{RPS(n)}
\label{RPS*}
\end{equation}

into a new quantity $RPS^*$ that we will consider when using our test-statistic.

The following pseudo code illustrates how the computation of the RPS value can be implemented:

\newcommand{\start}{\texttt{first}}
\newcommand{\last}{\texttt{last}}

\begin{algorithm}
\caption{Calculates the recursive product of spacings $rps$ from ordered samples $x_{(i)}$}
\begin{algorithmic} 

\State $x = [0, x_{(1)}, x_{(2)}, \dots, x_{(n)}, 1]$
\State $rps = 0$
    \State $s = x[\start +1:\last] - x[\start:\last-1]$ \Comment{initial spacings}
\While{$len(s) > 1$}
    \State $rps = rps - sum(log(s))$
    \State $s = s[\start:\last-1]+ s[\start+1:\last]$ \Comment{spacings for next iteration}
    \State $s = s/sum(s)$ \Comment{normalize}
\EndWhile
\State $normalized\_rps = min\_rps(n) / rps$
%\State{$rps /= min\_rps$}
\end{algorithmic}
\end{algorithm}

This algorithm has a computational complexity of $\mathcal{O}(n^2)$, and can become inefficient for very large sample sizes $n$. In this work we limit ourselves to $n\leq1000$.

In an analogue way, we can also define an extension to Greenwood $G(n)$, that instead of logarithms of spacings, sums over the squares of spacings. This means that we substitute  Eq.~\ref{M_j} with $M^{j} = \sum_{i=1}^{j}\left( s_i^j \right)^2$, while keeping the definition of $s_i^j$ from Eq.~\ref{sij}. We call this recursive form the "RSS" test statistic in the following comparison.

\subsection{Illustration}

\begin{figure}[H]
    \centering
    \includegraphics[width=0.8\textwidth]{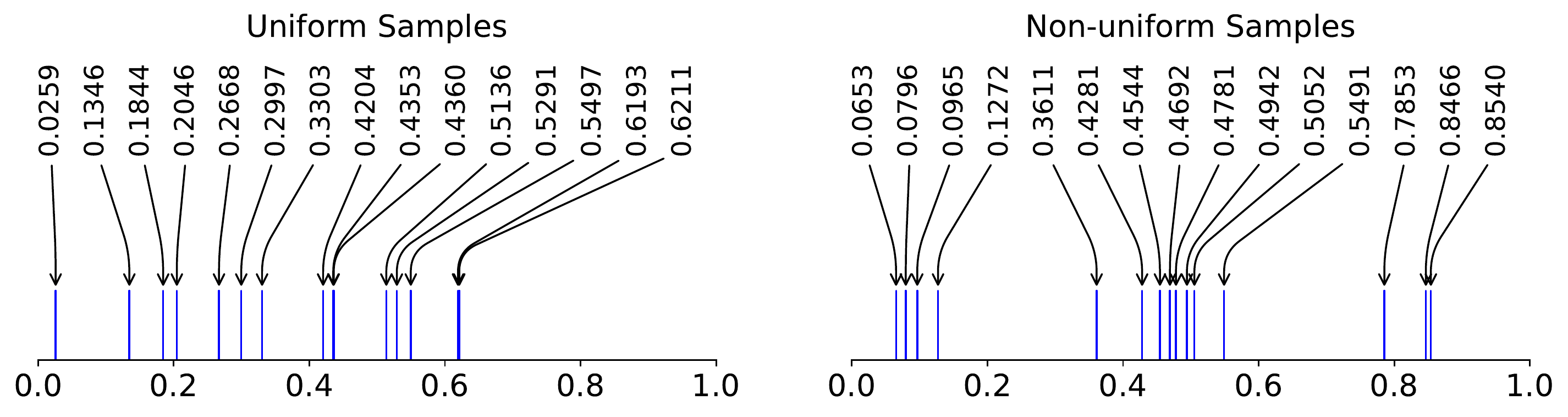}
    \caption{Example of 15 standard uniformly distributed samples (left) and 10 standard uniformly + 5 normally ($\mu=0.5, \sigma=0.1$) distributed samples (right). The sample positions on the $[0,1]$ interval are annotated via the arrows + text.}
    \label{fig:samples}
\end{figure}

To illustrate better how our test statistic works, and to highlight differences to other tests, we use the example set of samples drawn from a uniform (null hypothesis $H^0$) and a non-uniform distribution, respectively, shown in Fig.~\ref{fig:samples}.

The Moran test is based on the spacings between samples, and the smallest and largest spacings in the specific example are present in the uniform case. This leads to a more extreme test statistic value $t$ and hence p-value $p = P(T\geq t | H^0)$ of 0.117 for the uniform case, while it evaluates to $p=0.335$ in the non-uniform case.
The feature of samples clustering locally together, as in the non-uniform case, is---by construction---completely missed by Moran's test.

The KS test can detect such clustering via the CDF, however in our chosen example it is challenged by the fact that samples trend towards the left in the uniform case, while they are more balanced in the non-uniform case. This leads to p-values of 0.048 for uniform, and 0.356 for non-uniform, respectively.

\begin{figure}[H]
    \centering
    \includegraphics[width=0.8\textwidth]{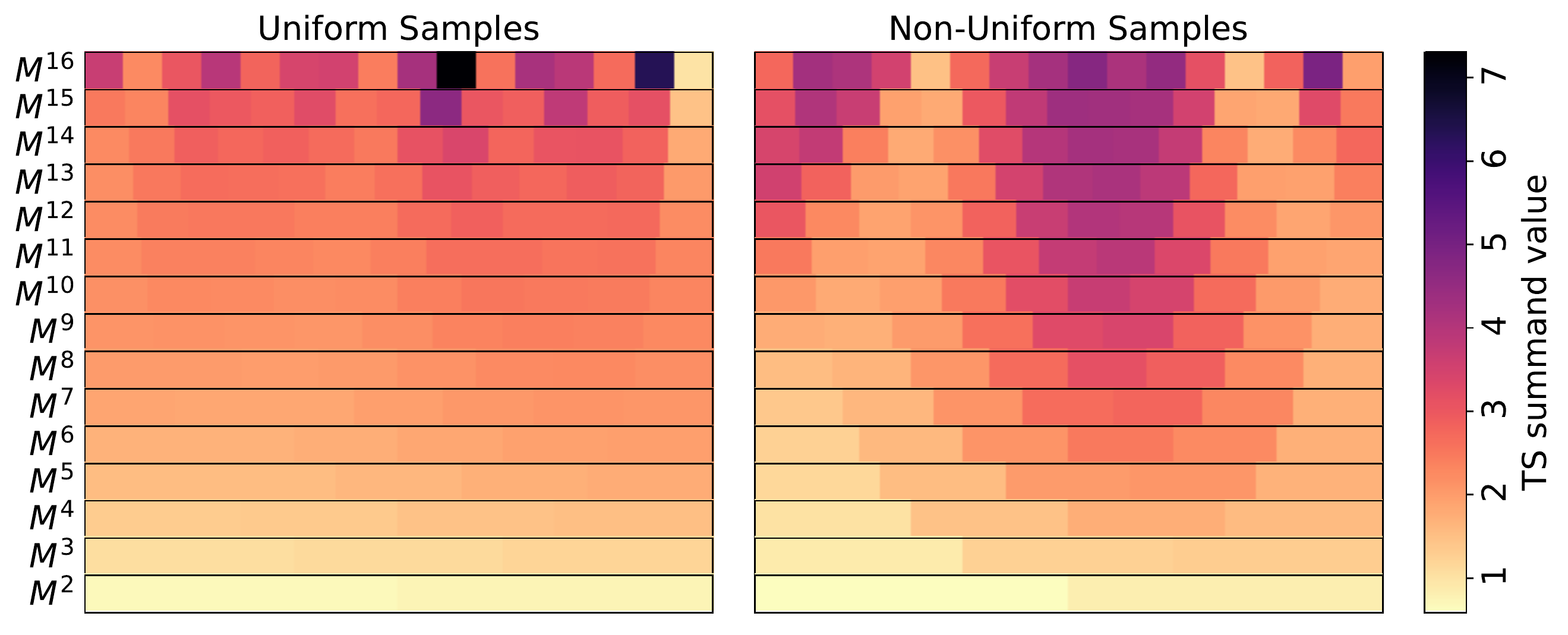}
    \caption{Illustration of the test statistic contributions from all recursion levels for the uniformly distributed samples (left) and the non-uniform samples (right). The sum over the first level only ($M^{16}$) is equivalent to the Moran statistic.}
    \label{fig:illustration}
\end{figure}

The RPS test, however, taking into account also spacing between spacings, finds a p-value of 0.532 for the uniform case, and a much lower p-value of 0.057 for the non-uniform samples.
The behaviour of RPS is further illustrated in Fig.~\ref{fig:illustration}, that shows the individual contribution of spacings of all recursion levels that build up the test statistic value. The Moran statistic corresponds to the sum over the first row ($M^{16}$), while all subsequent levels are added for RPS. The uniform samples exhibit the most extreme values in the first level, but then even out rapidly. Contrary to that, the non-uniform samples with their clustering give rise to larger values at later levels, explaining the lower observed p-value.

\section{Cumulative Distribution of RPS}
\label{sec:cdf}

In order to use RPS as a statistical test yielding p-values, we need its cumulative distribution $F$. In the case of $n=1$ that has only two spacings---the simplest non-trivial case we can encounter---we can easily derive the distribution of $RPS^*(1)$, which is:

\begin{equation}
    F_{RPS^*}(x;n=1) = 1 - \sqrt{1 - 4 ^{\frac{x-1}{x}}}
\label{eq:cdf_RPS_N_1}
\end{equation}

For $n\geq2$, however, it is not simple to derive this distribution. Therefore, we resort to numerically approximating the distribution of $RPS^*$ discussed in the following section. 

\subsection{Approximate Distribution}
\label{sec:rps_dist}

\begin{figure}[H]
    \centering
    \includegraphics[width=0.6\textwidth]{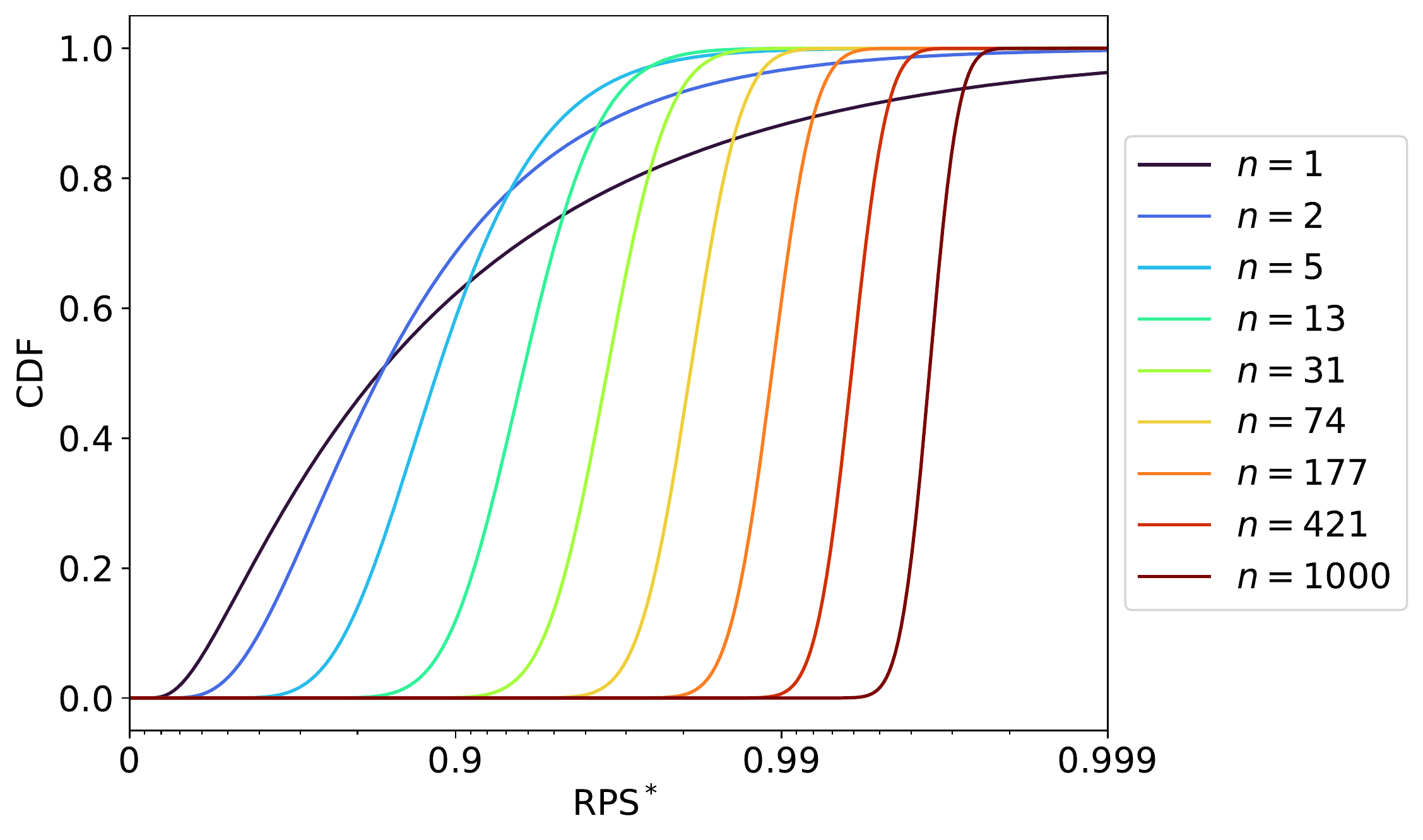}
    \caption{Example of CDFs of the $RPS^*$ distribution for a few different values of $n$. N.B.: the x-axis is displayed in \textit{inverted} logarithm.}
    \label{fig:cdfs}
\end{figure}

We have built an approximation for the cumulative distribution  $F_{RPS^*}(x;n)$ precise enough to compute meaningful p-values up to relatively extreme values of up to $10^{-7}$, and large sample sizes $n$ of up to 1000. Figure~\ref{fig:cdfs} shows some examples of $RPS^*$ distributions for a few values of $n$.

We base our approximation on simulation, drawing events with uniform distribution in the range $[0, 1]$ for a given $n$, and collecting $N=2 \cdot 10^8$ samples of $RPS^*(n)$.
Such simulation could be directly used to calculate p-value estimates by counting the fraction of trials below or above an observed $RPS^*$ value $x$ for a fixed $n$.
However, we want to provide a continuous and smooth function valid for any $n\leq1000$.
For this, we use simulated data to infer the values $x$ of our test statistics corresponding to a discrete list of specific quantiles $p \in [10^{-7} , 1 - 10^{-7}]$.
Taking the $i$-th element in the sorted simulation set gives an estimate for the value of $x(p=i/N)$.
In order to improve this estimate, we could use bootstrapping \citep{Efron:1979bxm}, collecting different realisations of $x$ by resampling the original dataset with replacement, resulting in a distribution of values of $x$ for each $p$, from which we can then extract the mean and the standard deviation, indicative of the error (see Fig~\ref{fig:x_N_fit}).
Instead of manually performing the bootstrapping, we can calculate the probability of each sample $x$ to represent a specific quantile $p$ if we were to sample randomly with replacement.
For simplicity, let us consider rational quantiles that can be expressed in the form $p=\frac{k}{N}$; the probability that the $i$-th sample could end up representing the $k$-th quantile is:

\begin{equation}
    \pi_{k,i} = F_B \left( k, N+1-k; \frac{i}{N} \right) - F_B \left( k, N+1-k; \frac{i-1}{N} \right)
\label{eq:probability_analytic_bootstrappig}
\end{equation}

where $F_B(a, b; t)$ is the cumulative function of the Beta distribution with parameters [$a$, $b$] estimated at $t$.
The distribution $\mathrm{Beta}(k, N+1-k)$ represents the $k$-th order statistic  of the uniform distribution \citep{DavidNagaraja:2003}, i.e. the $k$-th largest element of a set on $N$ uniformly distributed random variable.
Eq.~\ref{eq:probability_analytic_bootstrappig} corresponds to the limiting case of performing an infinite number of bootstrapping steps and can be used to quickly estimate the mean and standard deviation of all $x(p)$ for a choice on $n$, especially when dealing with large datasets:

\begin{equation}
    \mathrm{E} \left[ x\left( \frac{k}{N} \right) \right] = \sum_{i=1}^{N} x_i \cdot \pi_{k,i}
\label{eq:mean_analytic_bootstrapping}
\end{equation}

\begin{equation}
    \mathrm{Std} \left[ x\left( \frac{k}{N} \right) \right] =\sqrt{ \sum_{i=1}^{N} \left( x_i - \mathrm{E} \left[ x\left( \frac{k}{N} \right) \right] \right) ^2 \cdot \pi_{k,i} }
\label{eq:std_analytic_bootstrapping}
\end{equation}

It would be inefficient to produce such simulation for any $n$, and hence we repeat the above procedure for only 180 different choices of $n$ between 2 and 1000 following approximately a logarithmic spacing.

\subsection{Fitting procedure}
\label{sec:fitting_procedure}

Using Eq.~\ref{eq:mean_analytic_bootstrapping} and Eq.~\ref{eq:std_analytic_bootstrapping} we are able to define a grid of points with mean $\mu(n,p)$ and standard deviation $\sigma(n,p)$. Our goal is to estimate a set of points $\hat{x}(n,p)$, which will be the basis to interpolate and infer the distribution of the test statistic for all values of $n$ and $p$ defined above. The points $\hat{x}(n,p)$ is allowed to deviate from the means $\mu(n,p)$ within the uncertainties $\sigma(n,p)$, and can thereby provide a more accurate approximation by smoothing out stochastic noise. Additionally, points from the analytic solution for $n=1$ (Eq.~\ref{eq:cdf_RPS_N_1}) are added to the list as anchor points at the boundary.

\begin{figure}[H]
    \centering
    \includegraphics[width=0.8\textwidth]{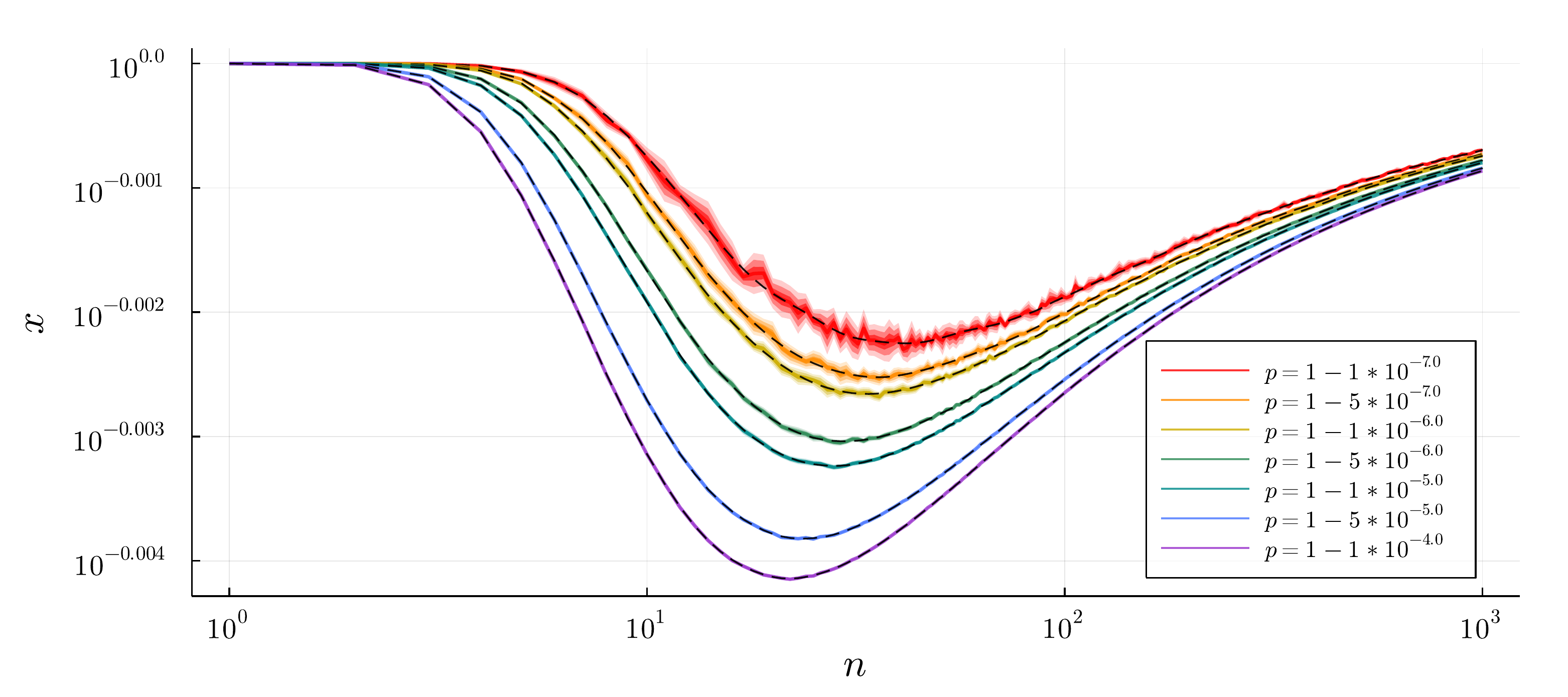}
    \caption{Example of spline fitted $x$-values across $n$ for a few extreme p-values. The colored bands show the 1, 2 and 3 sigma bands estimated via bootstrapping, the black, dashed lines show the approximations by the spline fits.}
    \label{fig:x_N_fit}
\end{figure}

Given a trial set $\tilde{x}(n,p)$, we interpolate a cubic spline polynomial across the values of $n$ for each value of $p$, similarly to the fits shown in Fig~\ref{fig:x_N_fit}. Given one such cubic spline, we evaluate the third derivative on both sides of each node, calculating the square of their difference and summing up across all nodes. Since we are using cubic splines, the third derivative is not continuous, and the "size" of the discontinuity is indicative of the smoothness of the interpolation. Summing up the contributions form all nodes of all cubic splines construct the smoothing cost function. The construction of this cost function is based on \citep{DIERCKX1975165, doi:10.1137/0719093, Dierckx1996CurveAS}, where smoothness is treated very similarly. The estimation of the cubic spline coefficients and the evaluation of the smoothness cost function can be represented as a quadratic objective function, which we want to minimize: 

\begin{equation}
    G(\tilde{x}) \propto \frac{1}{2} \tilde{x}^T \cdot Q \cdot \tilde{x} + \bar{h}^T \cdot \tilde{x} 
    \label{eq:smooth_cost_function}
\end{equation}

In addition to obtaining a smooth fit, there are also some additional constraints that need to be considered: monotonicity and sum of squared residuals.

Since the samples $\tilde{x}(p|n)$ should represent a cumulative density function, then it is important they are properly ordered, ensuring that $\tilde{x}(p_i|n) \leq \tilde{x}(p_j|n)$ for $i \leq j$. This is ensured including a number of linear inequality constraints modelled as a linear constraint matrix:

\begin{equation}
    A \cdot \tilde{x} \leq b
    \label{eq:monotonic_constraint}
\end{equation}

Lastly, we assume that the values $\tilde{x}(n,p)$ are normally distributed with means $\mu(n,p)$ and standard deviations $\sigma(n,p)$. Since we want to move away from the initial values $\mu(n,p)$ in order to obtain a smoother fit, it is important to limit this movement the further away we get and we do so by considering the sum of squared residuals, which is a typical measure to account for the global deviation from the mean. Since we assume gaussian deviations, the sum of all squared residuals can be modelled by a $\chi^2$ distribution with $m$ degrees of freedoms, where $m $ is the total number of parameters, i.e. the number of nodes. Given this distribution, we can estimate the value of the cost function to be limited to the mean ($m$) plus one standard deviation ($\sqrt{2m}$) of the $\chi^2$ distribution, thus:

\begin{equation}
    \sum_{i = 1}^m \frac{\left( \tilde{x}_i - \mu_i \right)^2}{\sigma_i^2} \leq m + \sqrt{2 \cdot m}
    \label{eq:chi2_constraint}
\end{equation}

Figure~\ref{fig:x_N_fit} shows a fitted spline representation of $\hat{x}(n|p)$ for different values of $p$.
Based on the resulting list of corresponding $p$ and $\hat{x}$ values, that we obtained for any $n$, we generate another spline interpolation as the approximation of the desired cumulative distribution $F(\hat{x}; n)$ for a given $n$.
As the cumulative distribution function $F$ is strictly monotonous in $\hat{x}$, we use the \cite{doi:10.1137/0905021} monotonic spline interpolation on the points $[\hat{x}(p|n), p]$ to produce the final CDFs, shown in Figure~\ref{fig:cdfs} for a few values of $n$. 

\subsection{Error estimation}
\label{sec:error_estimation}

\begin{figure}[h]
    \centering
    \begin{subfigure}[t]{0.49\textwidth}            \centering
        \includegraphics[width=\textwidth]{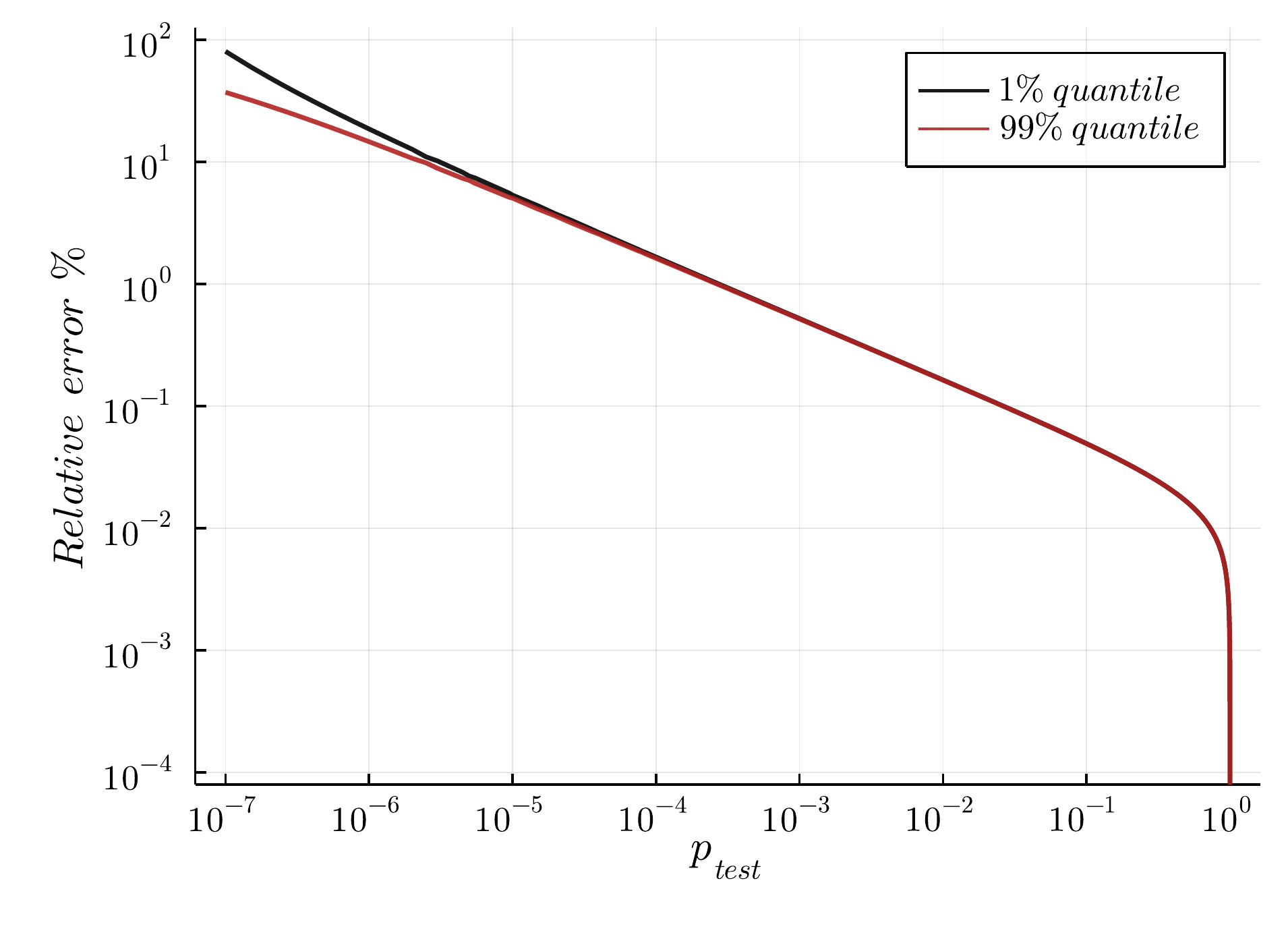}
        \caption{Estimated relative error of empirical p-value with respect to the 98 \% credible interval and $2 \cdot 10^8$ samples. The vertical axis reports the scale of the relative error in percent for two extremes, the 1\% and the 99\% quantile of the order statistic distribution.}
        \label{fig:analytic_error}
    \end{subfigure}
    \hfill
    \begin{subfigure}[t]{0.49\textwidth}  
        \centering 
        \includegraphics[width=\textwidth]{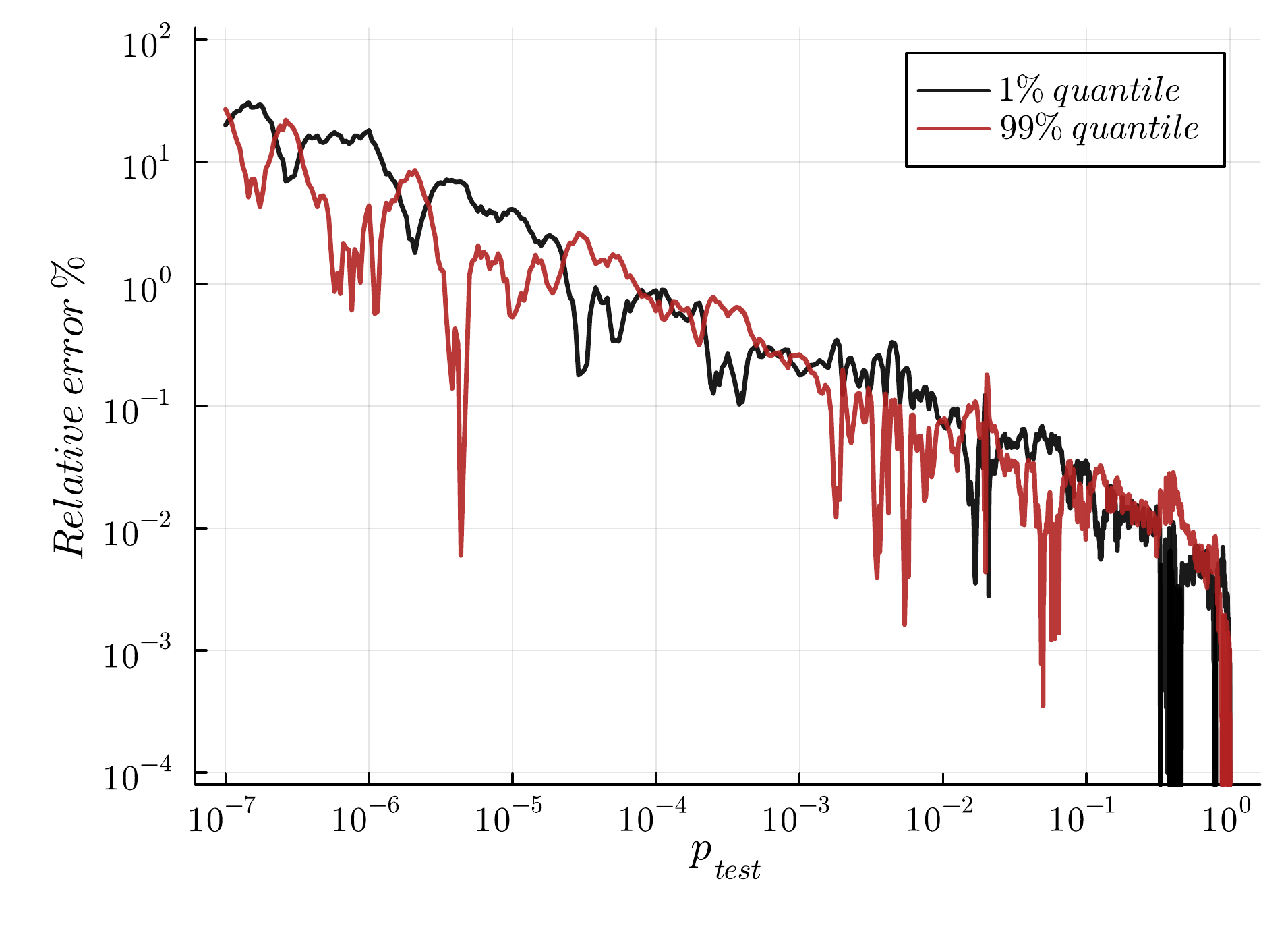}
        \caption{Estimated relative error of fitted p-value with respect to p-values obtained via bootstrapping. The vertical axis reports the scale of the relative error in percent for two extremes, the 1\% and the 99\% quantile of the bootstrapping distribution. Results for $n=75$.}
        \label{fig:numerical_error}
    \end{subfigure}
    \caption{Per-quantile relative error stimation of the approximate RPS distribution.}
    \label{fig:error_estimation}
\end{figure}

Finally, we are also able to estimate the precision of our approximation. Given any set of i.i.d. random variables, such as $x$, the corresponding list of estimated quantiles $p$ represents a random set of uniform variates. For any rational quantile $p_{test}=\frac{k}{N}$ we can estimate the 98\% credible interval $\left( p_{0.01}, p_{0.99} \right)$ using the distribution of the $k$-th order statistic $\mathrm{Beta}(k, N+1-k)$. Given the credible interval, we calculate the relative error of $p_{test}$ against the extrema of the interval, considering the largest value representative of the relative error of a random EDF up to a specified credible level. The results of the estimated relative error for our choice of $N=2 \cdot 10^8$ and for quantiles as low as $p=10^{-7}$ are shown in Figure~\ref{fig:analytic_error}.

As expected, the errors are increasing towards smaller p-values and exhibit an approximately linear behaviour in the log-log plot. We see that the estimated upper bound of the relative error for a p-value of $10^{-3}$ is below $1\%$, while for a p-value of $10^{-5}$ it increases to $<10\%$ and ultimately to $<100\%$ for p-values of $10^{-7}$. Such a "large" relative error for small p-values may sound alarming at first, but estimating a p-value of $10^{-7}$ and knowing it could actually be closer to $2 \cdot 10^{-7}$ would hardly change the statistical interpretation of a result.

In order to show the validity of these results, we compute the relative error of our approximate distributions against a test dataset containing 10 times more samples using bootstrapping. We do so for a few choices of number of events $n$, and in Figure~\ref{fig:numerical_error} it can bee seen that the behavior of the relative error is in complete agreement with our analytic estimates of Figure~\ref{fig:analytic_error}.

So defined, the relative error $\delta(p|N)$ is a function of the quantile $p$ and number of samples $N$, but this relationship can also be inverted in order to determine the number of samples necessary to achieve a desired relative error for a specific quantile: $N(p|\delta)$. Our choice of $N=2 \cdot 10^8$ was in fact guided by the requirement of having a relative error lower than 100\% for a p-value of $10^{-7}$ in at least 99\% of cases.

It is worth stressing that these estimates of the relative error are accurate with respect to the EDF that was sampled for each independent $n$, but might be subject to small changes after the smoothing fit we performed in order to regularize and infer the distributions for all missing values of $n$.

\subsection{Implementation}

The RPS test is made available as open-source packages for Python\footnote{https://pypi.org/project/spacings/} and Julia\footnote{https://juliapackages.com/p/spacingstatistics}, respectively, with the p-value parametrizations initially available up to 1000 samples.

Below we give a minimal example to evaluate the RPS test for an array $x$ in both language implementations, with x being:
\begin{lstlisting}[language=Python]
x = [0.1, 0.4, 0.76]
\end{lstlisting}

The python library can be used like the following:

\begin{lstlisting}[language=Python]
>>> from spacings import rps
>>> rps(x, "uniform")
RPStestResult(statistic=0.9547378863245608, pvalue=0.8865399970192409)
\end{lstlisting}

and the Julia equivalent giving identical results in the following:

\begin{lstlisting}
>>> using SpacingsTests
>>> rps(x, Uniform())
(statistic=0.9547378863245608, pvalue=0.8865399970192409)
\end{lstlisting}

\section{Example Application 1: Bump Hunting}

\newcommand{\ns}{\ensuremath{\langle n_s \rangle}}
\newcommand{\nb}{\ensuremath{\langle n_b \rangle}}

In this section, we illustrate how the RPS test could be used in a physics scenario. 
We consider a detector that collects a number of events in an observable $x$, where $x$ could for example be the energy of an event, the detection time, or a reconstructed quantity like an invariant mass.
We expect some or all of the observed events to follow a known background distribution $f_B(x)$, but there may be an additional contribution of events from an unknown signal distribution $f_S(x)$---such as a rare, exotic particle decay with unknown mass. Hence we want to quantify the goodness-of-fit of the background only model to our data. A resulting low p-value could indicate the presence of events distributed according to an additional, unknown signal distribution.

\begin{figure}[h]
    \centering
    \includegraphics[width=0.65\textwidth]{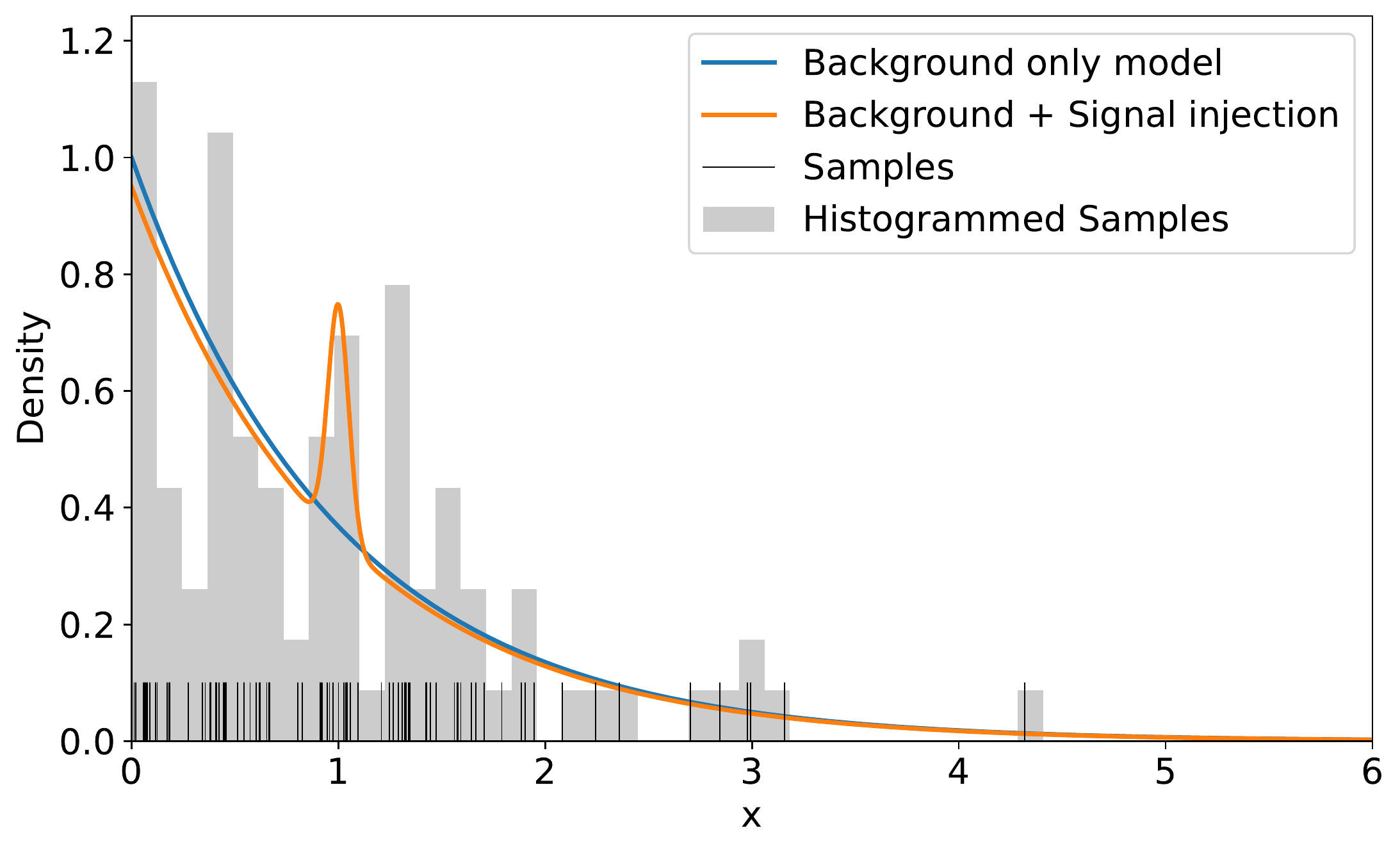}
    \caption{Example physics problem, with observed events distributed in $x$. We test the goodness-of-fit of the background only model (blue) to the samples. Here the samples have been generated according to a different distribution with an injected signal (orange).}
    \label{fig:example_problem}
\end{figure}

In the example here, we use an exponential distribution $f_B(x) = e^{-x}$ for the background model (null-hypothesis). In order to illustrate how the presence of an actual signal (alternative hypothesis) would affect the outcome, we also inject additional events following a normal distribution centred at $x=1$ and width $\sigma=0.05$.
The number of events is Poisson fluctuated for both background and signal, with expected values of $\nb=100$ and $\ns$ varied as specified.
In Fig.~\ref{fig:example_problem}, an example distribution of observed events is shown, together with the assumed background distribution, and the distribution with injected signal (here $\ns=5$).

The example case chosen is similar to that, for instance, of a search for an exotic particle with unknown mass ---a problem sometime referred to as "bump hunting". In this case, $x$ would represent an invariant mass.

N.B., we do not assume that we know the rate of the underlying processes, meaning that the number of observed counts is not included in our analysis other than for the calculation of the test statistic. This means that we test for the "shape" of the distribution, not its normalization. The conversion of events via the CDF of the distribution under test $f_B$ transforms the problem into a test of uniformity.

\begin{figure}[h]
    \centering
    \includegraphics[width=\textwidth]{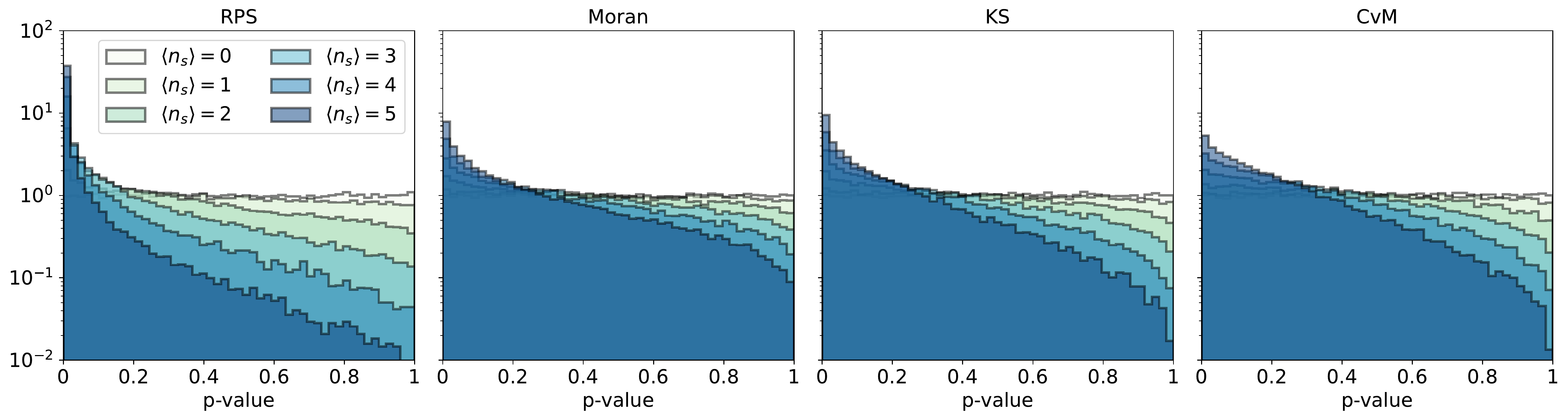}
    \caption{p-value distributions for background only samples ($\ns=0$) and background plus randomised signal injections comparing to the background model for several choices of test statistics.}
    \label{fig:example}
\end{figure}

The p-value distributions under the assumption of $H^0$ (i.e. only background is present) for repeated trials with $\nb=100$, and various injected $\ns= [0, 3, 6, 9, 12, 15]$ are shown in Fig.~\ref{fig:example}. All distributions with no signal ($\ns= 0$) show a flat p-value distribution as expected, since in that case all events are drawn from the background distribution $p_B$. For trials with injected signal, the distributions are trending towards smaller p-values, indicating the worsened goodness-of-fit for the background only model. In the example, all tests exhibit this behaviour, while the RPS test offers the largest rejection probability of the null hypothesis.

\begin{figure}[h]
    \centering
    \includegraphics[width=0.5\textwidth]{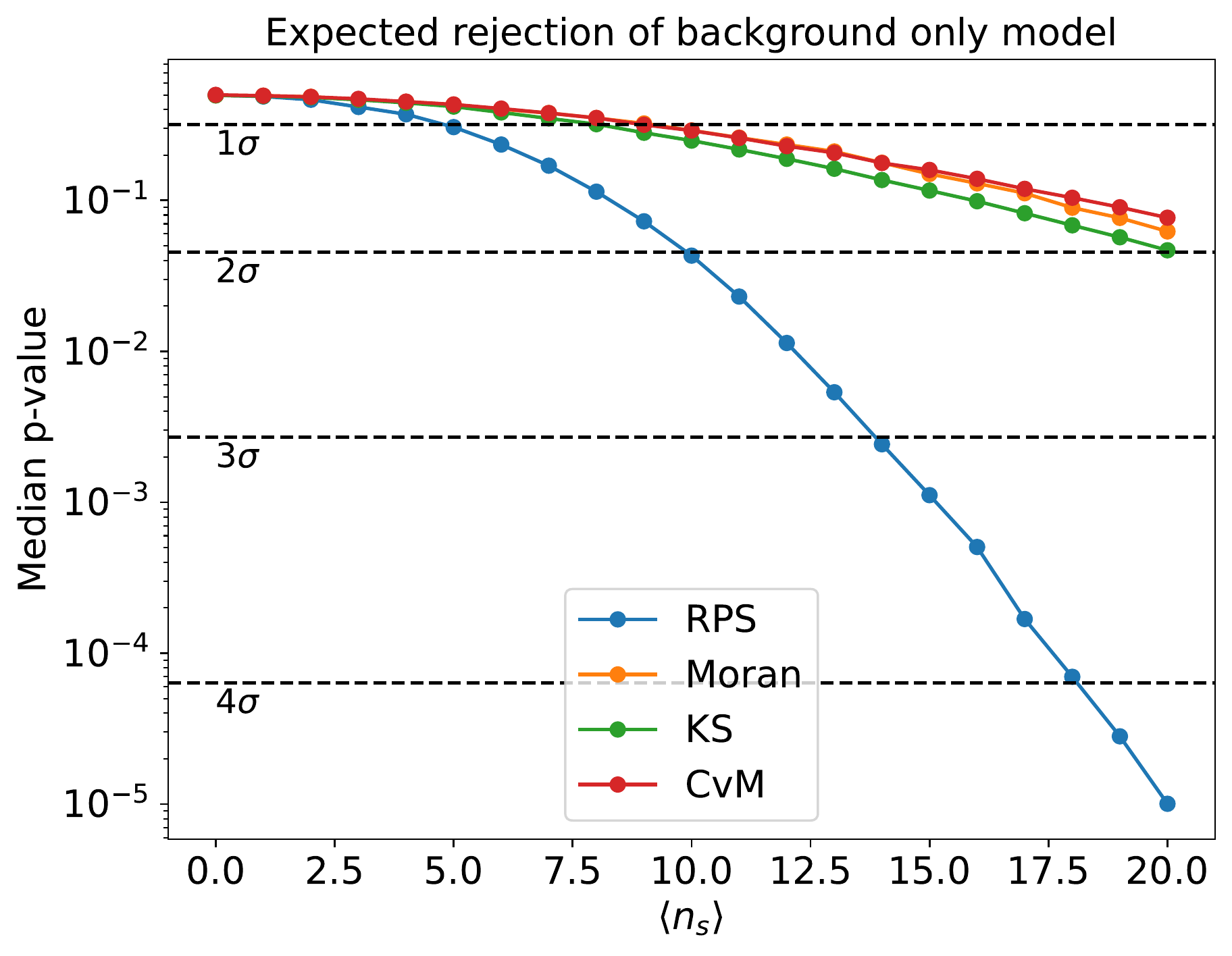}
    \caption{The expected significance level at which the background model can be excluded under the assumption of a signal, as a function of \ns\ for the different tests. }
    \label{fig:sensitivity}
\end{figure}

We quantify the sensitivity of the analysis to reject the background only model at different significance levels under the assumption of the presence of a signal.
Therefore we check the median p-value of repeated trials, and at what value of $\ns$ it crosses specific critical values (See left panel of Fig.~\ref{fig:sensitivity}).
In our chosen example, for a signal of strength $\ns=10$ we expect to reject the background only model using RPS at the $2\sigma$ significance level\footnote{A significance level in terms of numbers of $k$ standard deviations $\sigma$ can be translated to a p-value as one minus the integral over a unit normal distribution form $-k$ to $+k$.}, whereas for the other tests, a signal of at least $\ns=20$ is needed to achieve the same. Such a large signal of $\ns=20$ would allow to reject the background only model at $> 4\sigma$ significance with the RPS test.

\section{Example Application 2: Trigger for Transient Neutrino Emission}

This section summarizes one of the first applications of the RPS test in astrophysics, namely for triggering transient events in cryogenic neutrino detectors~\citep{Eller:2022ddl} such as the RES-NOVA experiment~\cite{RES-NOVA:2021gqp}.
While the technical details about the experimental setup, the simulation and the application of the RPS test can be found in the aforementioned references, here we will summarize some highlights.

\begin{figure}[h]
    \centering
    \includegraphics[width=0.8\textwidth]{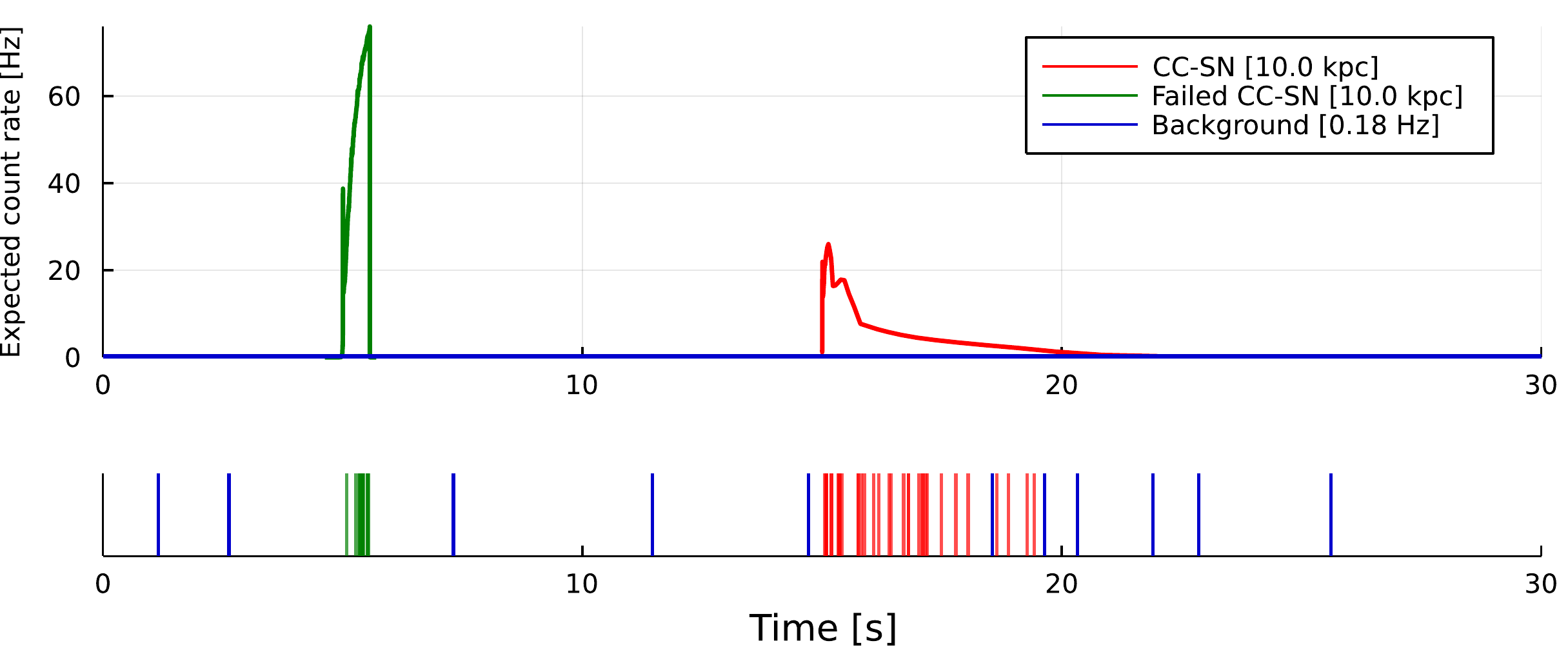}
    \caption{Example of observed counts at a neutrino detector for signals from a core-collapse SN (at time $t=15$\,s) and a failed core-collapse SN (at $t=5$\,s) for progenitors stars with 27~$M_{\odot}$ and 40~$M_{\odot}$ respectively, both at a distance of 10\,kpc. (Modified version of a Figure from Ref.~\citep{Eller:2022ddl})
    }
    \label{fig:sn}
\end{figure}

Cryogenic neutrino detectors can be described as counting experiments, that output a temporal data stream of observed neutrino interactions. Without the presence of a transient neutrino source, we only expect some activity from background events. If a source of neutrinos is placed at an observable distance, such as a core-collapse supernova (CC-SN) at 10\,kpc, we expect a short burst of neutrinos resulting in an excess in the observed counts over the background only expectation.
The sources of such transient neutrinos can vary in their overall duration, temporal distribution and amplitude. Figure~\ref{fig:sn} shows as examples the expected counts of two different neutrino sources, a CC-SN and a failed CC-SN, respectively, together with a constant background expectation at a rate of 0.18\,Hz.

\begin{figure}[h]
    \centering
     \includegraphics[width=0.8\textwidth]{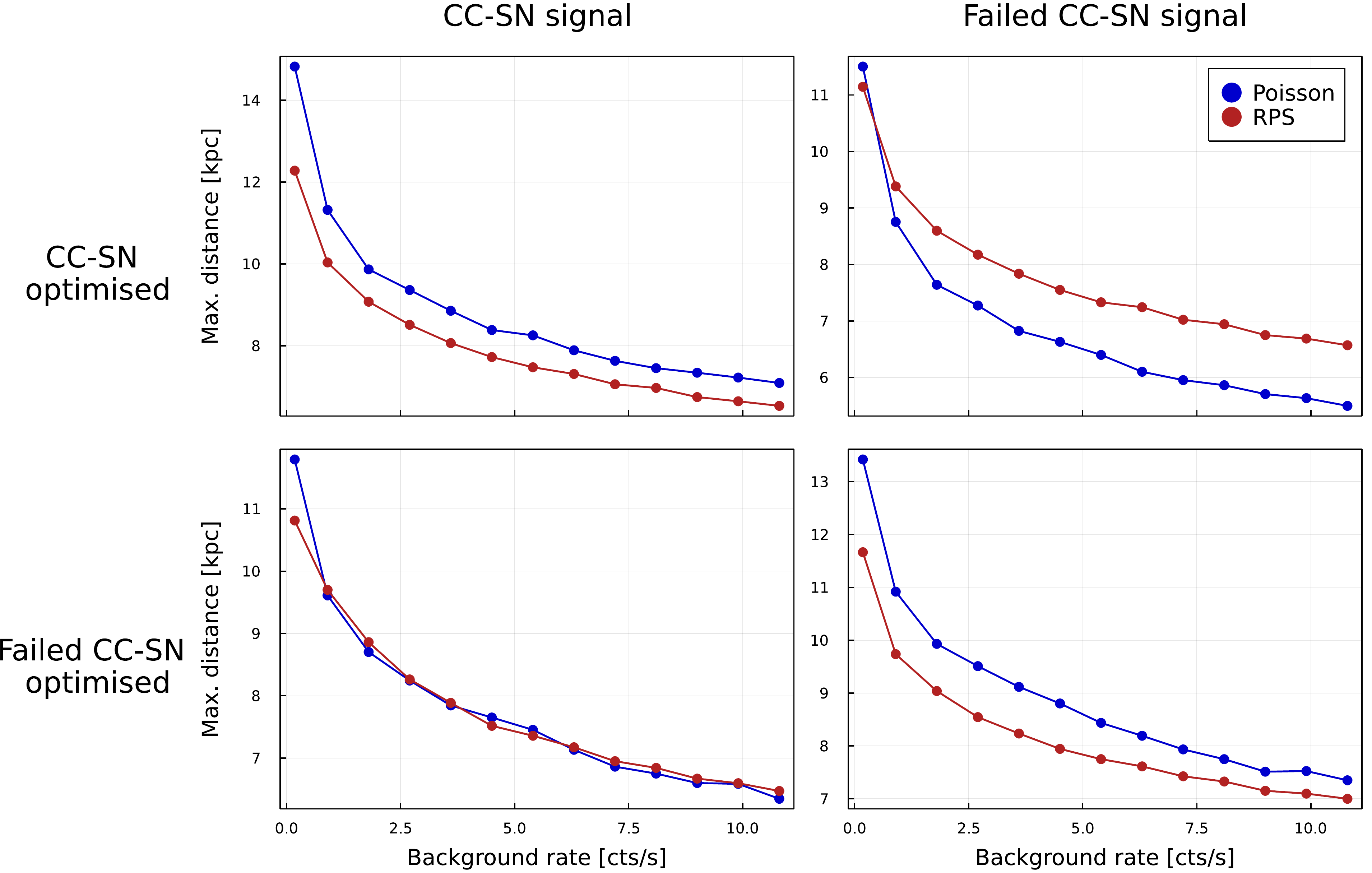}
    \caption{Maximum distance probed at a 95\% success rate as a function of time with respect to two sample signals, the Core-Collapse SN and the failed Core-Collapse SN, obtained using analysis windows optimised on each of the tested signals. (Figure from Ref.~\citep{Eller:2022ddl})}
    \label{fig:SN_and_BH_optimised_windows}
\end{figure}

To issue alerts in near real time about the presence of such sources, one needs a triggering system with a chosen false alarm rate (FAR), which is set to 1 per week for SNEWS~\cite{SNEWS:2020tbu}.
The standard approach for building such triggers is the usage of Poisson statistics, that analyse the data stream in windows of a fixed length, and check the level of observed counts compared to the expectation from background, see for example Ref.~\cite{Agafonova:2007hn}.
The Poisson approach works well if the window size is chose optimally for a given signal. However, if the chosen window size does not match the signal, the performance is affected as either the signal is not contained in the window (window too small), or the window is too large and the signal is washed out by background events. Performance of the Poisson test, as a function of the background rate and for the case of optimal window choice for the CC-SN and the failed CC-SN signals is shown in Fig.~\ref{fig:SN_and_BH_optimised_windows}.

The RPS test can likewise be used to analyze data streams to look for transient phenomena. Here we do not make any explicit assumption on the background rate, but rather assume that the background is constant in rate, which means the distribution of background events in the time dimension is following a uniform distribution. With RPS we can test for this uniformity, which can be used to detect short additional contribution of events in the data. The performance of the RPS test as trigger is also shown in Fig.~\ref{fig:SN_and_BH_optimised_windows}.

In the case where the window size for the Poisson test is optimal, the performance can not be matched with RPS (the panels in the upper left and lower right, respectively, in Fig.~\ref{fig:SN_and_BH_optimised_windows}) and results in up to 10\% lower sensitivity.
However, the more interesting case is when using the window size optimized for one signal for the analysis of a different signal (the panels in the upper right and lower left, respectively, in Fig.~\ref{fig:SN_and_BH_optimised_windows}). The RPS test is more robust to such changes, and in the example of searching for a failed CC-SN signal with a window optimized on a particular CC-SN scenario, we find up to 20\% increase in sensitivity.

In general, what we find is that the RPS test being non parametric and able to deal with much larger analysis windows is more robust to changing conditions. Less assumptions about the background rate and the expect signals have to be made at the trade off of being non-optimal to one specific signal choice, but good performance for the more agnostic case of unknown signal distributions. This makes RPS an interesting choice for a general-purpose, agnostic trigger algorithm for the search of transient events.

\section{Performance Comparison}

This section presents an in-depth performance comparison of the RPS test to several other tests referenced in the introduction (KS, AD, CvM and Moran---all those that allow to compute p-values). We are interested in detecting small changes in an otherwise uniform distribution, and therefore construct the following generic benchmark scenario: For one simulation of a specific test case $H^K(n, s, w)$ we generate $(1-s) \cdot n$ random variates\footnote{Numbers of samples are rounded to the closest integer} from a standard uniform distribution $\mathcal{U}(0,1)$, where $s$ is a \textit{signal} fraction. In addition, we include $s\cdot n$ samples distributed according to $\Delta + \mathcal{U}(0,w)$ with the offset $\Delta = \mathcal{U}(0,1-w)$, i.e. a more narrow uniform distribution of width $w$ over a random interval within $(0,1)$.
In our comparison, we vary all three parameters of $H^K(n, s, w)$, i.e. the number of samples $n$, as well as the fraction $s$ and width $w$ of the injected \textit{signal} events.
A sensitive test should be able to detect the presence of the added, narrower signal samples by reporting a low p-value.

\begin{figure}[h!]
    \centering
    \includegraphics[width=\textwidth]{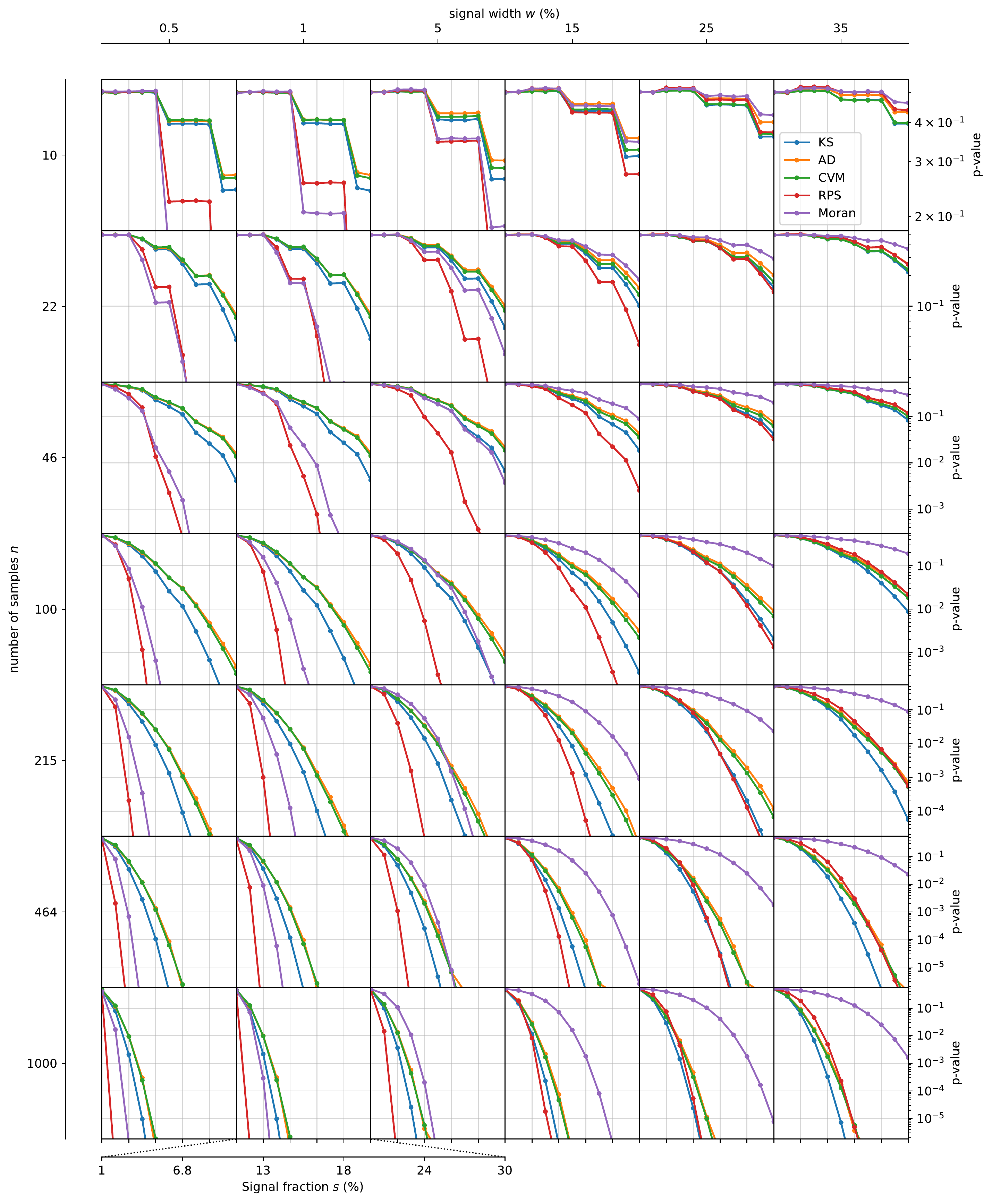}
    \caption{Comparison of the performance (median p-value of repeated trials, individual panel's xy-axis) as a function of the number of total samples (large y-axis), the width of the signal (large x-axis) samples, and the fraction of signal samples (individual panel's x-axis). The number of signal samples is rounded to the closest integer, hence the "step"-like features visible mostly in the first few rows.}
    \label{fig:comparison}
\end{figure}

Figure~\ref{fig:comparison} show the performance of our choice of tests as a function of the above three parameters.
As a metric, we show the median p-value obtained from repeated trials, and we interpret a lower reported median p-value as a more powerful test. This number can be interpreted as the median significance at which we expect to be able to reject the null hypothesis.
What can be observed is, that for all the tested scenarios the RPS test is performing either on par or significantly better than the Moran test. The EDF based tests (KS, AD or CvM) start to dominate in terms of performance only for relatively wide signals of around 25\% total width or more.
When analysing the goodness-of-fit given a large number of samples, i.e. order of several hundreds, the differences between RPS and the EDF-style tests start to become smaller.
Overall, the outcome of this performance study suggests that when signals are expected of widths that span over less than a 25\% percentile of the null hypothesis distribution, and if the number of samples is $n<1000$, the RPS tests compares very favourably against all others considered.

We also investigated other metrics to judge the test's performance, such as the area under the receiver operating characteristics (ROC) curve  between signal and null hypothesis trials. The overall picture does not change substantially.

\section{Conclusions}

The RPS test statistic is a sensitive measure to detect deviations of samples from a continuous distribution with known CDF.
The analytic distribution of the RPS statistic is not available for $n>1$, but a high accuracy parameterization valid up to sample sizes of $n=1000$ is provided in order to use RPS as a goodness-of-fit test.
In the presented test scenarios, the RPS test outperforms other tests significantly under certain circumstances, in particular when the observed sample is small ($n<1000$) and introduced deviations are narrow, i.e. concentrated over a small quantile.
Two example physics analysis cases were presented, we show that the sensitivity of a "bump hunting" experiment could be boosted by up to a factor of two by choosing the RPS test over others. And we show how RPS can be used to build a robust and agnostic trigger algorithm for a SN experiment.

\section*{Acknowledgements}

We would like to thank Allen Caldwell, Oliver Schulz and Johannes Buchner for helpful discussions and comments.\\
This research was supported by the Deutsche Forschungsgemeinschaft (DFG, German Research Foundation) under Germany´s Excellence Strategy – EXC-2094 – 390783311

\bibliography{references}  

\providecommand{\href}[2]{#2}\begingroup\raggedright\begin{thebibliography}{10}

\bibitem{Kolmogorov}
A.~Kolmogorov, \emph{Sulla determinazione empirica di una legge di
  distribuzione}, {\emph{G. Ist. Ital.} {\bfseries 4} (1933) }.

\bibitem{Smirnov1948TableFE}
N.~Smirnov, \emph{Table for estimating the goodness of fit of empirical
  distributions}, {\emph{Annals of Mathematical Statistics} {\bfseries 19}
  (1948) 279}.

\bibitem{AD}
T.W.~Anderson and D.A.~Darling, \emph{A test of goodness of fit},
  \href{https://doi.org/10.1080/01621459.1954.10501232}{\emph{Journal of the
  American Statistical Association} {\bfseries 50} (1954) 765}.

\bibitem{Marhuenda}
Y.~Marhuenda, D.~Morales and M.C.~Pardo, \emph{A comparison of uniformity
  tests}, \href{https://doi.org/10.1080/02331880500178562}{\emph{Statistics}
  {\bfseries 39} (2005) 315}.

\bibitem{cramer}
H.~Cramer, \emph{On the composition of elementary errors},
  \href{https://doi.org/10.1080/03461238.1928.10416862}{\emph{Scandinavian
  Actuarial Journal} {\bfseries 1928} (1928) 13}.

\bibitem{VonMises}
R.V.~Mises, \emph{Wahrscheinlichkeit Statistik und Wahrheit}, Springer-Verlag
  Berlin Heidelberg (1928),
  \href{https://doi.org/10.1007/978-3-662-36230-3}{10.1007/978-3-662-36230-3}.

\bibitem{Pyke}
R.~Pyke, \emph{{The Supremum and Infimum of the Poisson Process}},
  \href{https://doi.org/10.1214/aoms/1177706269}{\emph{The Annals of
  Mathematical Statistics} {\bfseries 30} (1959) 568 }.

\bibitem{Durbin}
J.~Durbin, \emph{Tests for serial correlation in regression analysis based on
  the periodogram of least-squares residuals}, {\emph{Biometrika} {\bfseries
  56} (1969) 1}.

\bibitem{Brunk}
H.D.~Brunk, \emph{{On the Range of the Difference between Hypothetical
  Distribution Function and Pyke's Modified Empirical Distribution Function}},
  \href{https://doi.org/10.1214/aoms/1177704578}{\emph{The Annals of
  Mathematical Statistics} {\bfseries 33} (1962) 525 }.

\bibitem{10.2307/2336673}
R.C.H.~Cheng and M.A.~Stephens, \emph{A goodness-of-fit test using moran's
  statistic with estimated parameters}, {\emph{Biometrika} {\bfseries 76}
  (1989) 385}.

\bibitem{Greenwood}
M.~Greenwood, \emph{The statistical study of infectious diseases},
  {\emph{Journal of the Royal Statistical Society} {\bfseries 109} (1946) 85}.

\bibitem{CressieL}
N.~Cressie, \emph{On the logarithms of high-order spacings}, {\emph{Biometrika}
  {\bfseries 63} (1976) 343}.

\bibitem{CressieS}
N.~Cressie, \emph{An optimal statistic based on higher order gaps},
  {\emph{Biometrika} {\bfseries 66} (1979) 619}.

\bibitem{shtembari2020sum}
L.~Shtembari and A.~Caldwell, \emph{On the sum of ordered spacings},  2020.

\bibitem{Efron:1979bxm}
B.~Efron, \emph{{Bootstrap Methods: Another Look at the Jackknife}},
  \href{https://doi.org/10.1214/aos/1176344552}{\emph{Annals Statist.}
  {\bfseries 7} (1979) 1}.

\bibitem{DavidNagaraja:2003}
H.A.~David and H.N.~Nagaraja, \emph{Order statistics}, Wiley (2003).

\bibitem{DIERCKX1975165}
P.~Dierckx, \emph{An algorithm for smoothing, differentiation and integration
  of experimental data using spline functions},
  \href{https://doi.org/https://doi.org/10.1016/0771-050X(75)90034-0}{\emph{Journal
  of Computational and Applied Mathematics} {\bfseries 1} (1975) 165}.

\bibitem{doi:10.1137/0719093}
P.~Dierckx, \emph{A fast algorithm for smoothing data on a rectangular grid
  while using spline functions},
  \href{https://doi.org/10.1137/0719093}{\emph{SIAM Journal on Numerical
  Analysis} {\bfseries 19} (1982) 1286}
  [\href{https://arxiv.org/abs/https://doi.org/10.1137/0719093}{{\ttfamily
  https://doi.org/10.1137/0719093}}].

\bibitem{Dierckx1996CurveAS}
P.~Dierckx, \emph{Curve and surface fitting with splines},  in \emph{Monographs
  on numerical analysis}, 1996.

\bibitem{doi:10.1137/0905021}
F.N.~Fritsch and J.~Butland, \emph{A method for constructing local monotone
  piecewise cubic interpolants},
  \href{https://doi.org/10.1137/0905021}{\emph{SIAM Journal on Scientific and
  Statistical Computing} {\bfseries 5} (1984) 300}
  [\href{https://arxiv.org/abs/https://doi.org/10.1137/0905021}{{\ttfamily
  https://doi.org/10.1137/0905021}}].

\bibitem{Eller:2022ddl}
P.~Eller, N.~Iachellini~Ferreiro, L.~Pattavina and L.~Shtembari, \emph{{Online
  triggers for supernova and pre-supernova neutrino detection with cryogenic
  detectors}},  \href{https://arxiv.org/abs/2205.03350}{{\ttfamily
  2205.03350}}.

\bibitem{RES-NOVA:2021gqp}
{\scshape RES-NOVA} collaboration, \emph{{RES-NOVA sensitivity to core-collapse
  and failed core-collapse supernova neutrinos}},
  \href{https://doi.org/10.1088/1475-7516/2021/10/064}{\emph{JCAP} {\bfseries
  10} (2021) 064} [\href{https://arxiv.org/abs/2103.08672}{{\ttfamily
  2103.08672}}].

\bibitem{SNEWS:2020tbu}
{\scshape SNEWS} collaboration, \emph{{SNEWS 2.0: a next-generation supernova
  early warning system for multi-messenger astronomy}},
  \href{https://doi.org/10.1088/1367-2630/abde33}{\emph{New J. Phys.}
  {\bfseries 23} (2021) 031201}
  [\href{https://arxiv.org/abs/2011.00035}{{\ttfamily 2011.00035}}].

\bibitem{Agafonova:2007hn}
N.Y.~Agafonova et~al., \emph{{On-line recognition of supernova neutrino bursts
  in the LVD detector}},
  \href{https://doi.org/10.1016/j.astropartphys.2007.09.005}{\emph{Astropart.
  Phys.} {\bfseries 28} (2008) 516}
  [\href{https://arxiv.org/abs/0710.0259}{{\ttfamily 0710.0259}}].

\end{thebibliography}\endgroup
\bibliographystyle{JHEP.bst}

\end{document}